# Multifractality in Surface Potential for Cancer Diagnosis

Phat K. Huynh [1], Dang Nguyen [2], Grace Binder [3], Sharad Ambardar [2], Trung Q. Le [1,2], and Dmitri V. Voronine[2,4*]

[1] Department of Industrial and Management Systems Engineering, University of South Florida, Tampa, FL, 33620

[2] Department of Medical Engineering, University of South Florida, Tampa, FL, 33620

[3] Department of Chemistry, University of South Florida, Tampa, FL, 33620

[4] Department of Physics, University of South Florida, Tampa, FL, 33620

* Corresponding author: dmitri.voronine@gmail.com

## Abstract

Recent advances in high-resolution biomedical imaging focusing on morphological, electrical, and biochemical properties of cells and tissues, scaling from cell clusters down to the molecular level, have improved cancer diagnosis. Multiscale imaging revealed high complexity that requires advanced data processing methods of multifractal analysis. We performed label-free multiscale imaging of surface potential variations in human ovarian and breast cancer cells using Kelvin probe force microscopy (KPFM). An improvement in the differentiation between normal and cancerous cells of for multifractal analysis using adaptive versus median threshold for image binarization was demonstrated. The results reveal the potential of using multifractality as a new biomarker for cancer diagnosis. Furthermore, the surface potential imaging can be used in combination with morphological imaging for cancer diagnosis.

# 1. INTRODUCTION

Breast and ovarian are the most frequent types of cancer among women worldwide [1, 2]. Morbidity and mortality of cancer are substantially decreased with early detection [3]. Cytological screening tests have decreased mortality. However, these methods have insufficient sensitivity and are time-consuming in both analysis and training of professionals with subjective manual diagnosis. More accurate tests may substantially decrease the cost and patient inconvenience. Tumorigenesis is a complex process with an uncontrolled growth of cells that ignore apoptotic signals triggered by cell cycle dysregulation and modulate cell survival pathway signaling. This process involves remodeling of the extracellular matrix, accompanied by morphological and electrochemical changes in the plasma membrane. Nanoscale imaging techniques investigate these changes with a high spatial resolution to better understand tumorigenic mechanisms.

The plasma membrane resting potential was shown to undergo abnormal depolarization in cancer cells [4, 5]. Various mechanisms of membrane potential regulation have been investigated that involve cell signaling pathways mediated by the disrupted activities of ion channels, pumps, and transporters. The potential difference between tumor and paratumor was found for several types of cancer, and the resulting depolarization was correlated with metastasis [6]. The membrane potential has been identified as an important bioelectric marker that reflects the changes in cellular activities.

The common methods of membrane potential measurements based on electrodes and voltage-sensitive dyes have a lack of imaging and a limited spatial resolution, respectively. KPFM is a nanoscale electrostatic force imaging technique based on the contact potential difference (CPD) between a scanning probe tip and sample [7], which has been less commonly used for cell imaging. KPFM has a high spatial resolution of less than 10 nm, which is determined by the size of the tip apex. It was previously used in a variety of biomedical applications such as mapping the surface potential of biomolecules [8-10], including DNA [11, 12] proteins [13, 14], and plasma membrane of cells [15, 16], revealing biomolecular interactions at the single-molecule level [8, 17-19].

Nanoscale morphological measurements by atomic force microscopy (AFM) have previously been used for cancer detection [19-21]. AFM provides direct imaging of cell surface morphology with nanoscale resolution. Physical properties such as cell stiffness, adhesion, and elasticity were used to identify cancerous tissue [22, 23]. However, in some cases AFM morphological imaging cannot differentiate between normal and cancer cells, while adhesion maps showed differences in fractality [24]. Fractal [25-30] and multifractal [31, 32] analyses were previously used for cancer diagnosis. Fractal geometry was used to describe the morphology of cancer cells and tissues by a single parameter, the fractal dimension, as a diagnostic biomarker [33, 34]. The change in fractal dimension or self-similar organization of surface morphology by malignant transformation can be quantified. Fractal dimension is suitable for the characterization of monofractal objects that have the same scaling exponent at different scales. However, a more complex organization (*e.g.*, cellular membrane) exhibits different fractal exponents at different scaling ranges resulting in several interwoven fractal sets, which are better described by the multifractal formalism [35, 36]. Fractal and multifractal analyses significantly improved the diagnostic efficiency of AFM imaging for cancer detection. However, the interpretation of complex morphologies was limited due to the lack of corresponding molecular information. KPFM is a nanoscale imaging technique based on molecular bioelectricity that can provide deeper insights.

In this work, we performed the fractal and multifractal analysis of morphological (AFM) and bioelectric (KPFM) images of ovarian and breast cells. The high spatial resolution allows for probing a broad range of scales, covering more than three orders of magnitude, from < 10 nm to tens of micrometers. We used the box-counting method to determine the fractal dimensions with significant variations at different scaling ranges. We used multifractal analysis for a more precise characterization of the scaling behavior, which showed a significant difference between the surface potential of the normal and cancer cells. We showed an improved efficiency of KPFM compared to AFM multifractality for cancer detection.

## 2. MATERIALS AND METHODS

### 2.1. Sample preparation

**SHT290 and ES-2 cell lines**

The human ovarian cancer ES-2 CCOC cell line (American Type Culture Collection, USA) [37] was cultured in MCDB 131: Media 199 (1:1 ratio) and McCoy's medium, respectively, supplemented with 10% FBS, 100 units/ml of penicillin and 100μg/ml of streptomycin. The immortalized normal human endometrial stromal cell line, SHT290 (Kerafast, USA) [38] was maintained in F12K : Media 199 (1:1 ratio) and supplemented with 5% FBS, 0.1% Mito, 2 μg/ml of human insulin, 100 units/ml of penicillin and 100 μg/ml of streptomycin. All cell cultures were passaged less than 10 times. Cell cultures were maintained in an incubator at 37°C and 5% $CO_2$ atmosphere.

**MCF-10A and MDA-MB-231 cell lines**

MCF-10A cells (Michigan Cancer Foundation) are a non-tumorigenic, immortalized mammary epithelial cell line, while MDA-MB-231 cells are triple-negative metastatic breast cancer epithelial cells [39-43] and were grown according to ATCC guidelines.

**Fixation**

All cells were cultured on atomically flat gold substrates (Tedpella) as required for KPFM measurements. Cells were washed with phosphate-buffered saline (PBS) before and after fixation to remove residual media or fixative solution, respectively. Samples were fixed using 4% paraformaldehyde (PFA) for 15 min at room temperature. After washing, fixed cells were stored in PBS with 0.01% sodium azide. Samples were rinsed with methanol prior to analysis.

**AFM and KPFM**

Both AFM and KPFM measurements were performed in tapping mode with an average tip-sample distance of 20 nm. AFM measurements were performed using Si tip, whereas KPFM measurements were performed using a conductive Au-coated Si tip. All images used for fractal and multifractal analysis were acquired with 3×3 $\mu m^2$ scan sizes, 512×512 data point resolution, and with a pixel size of ~ 6 nm. AFM and KPFM imaging were performed in air at ambient temperature (see schematic diagrams in supplementary Fig. S1).

***2.2. Image sampling and image binarization using adaptive and median thresholding***

Figure 1 shows the typical examples of the optical (Figs. 1a – 1d), AFM (Figs. 1e – 1h) and KPFM (Figs. 1i – 1l) images of the 37four cell lines used in this work. The cancer detection procedure involved the following steps. In step 1, the simultaneous AFM/KPFM image sampling was performed by scanning 3×3 $\mu m$ areas selected on the cytoplasmic and nucleus parts of the cells with 512×512 pixel density as described in the Methods section (Fig. 1m). After normalization, in step 2, the image binarization was performed using either the adaptive threshold or median methods (Fig. 1n). In step 3, either the fractal or multifractal analysis was performed (Fig. 1o) that delivered parameters for the statistical analysis.

To convert 2D gray-scale AFM/KPFM images to binary images, we used two image binarization methods: adaptive and median thresholding. The adaptive image thresholding technique is called Bradley's method [44], in which the method binarizes the gray-scale image using a locally adaptive threshold. The threshold was estimated for each pixel utilizing the local mean intensity around the neighborhood of the pixel. The adaptive method uses a neighborhood size of ~1/8[th] of the image size (64 pixels). If the current pixel value is less than the computed threshold, then it is set to black (i.e., value 0), otherwise, it is set to white (i.e., value 1). The median thresholding image binarization uses the median value of all pixels as a hard threshold and set the pixels whose values are larger than the median to black and set them to white otherwise.

### *2.3. Fractal and multifractal methods*

**Fractal analysis**

Fractal dimension is a measure of the space-filling capacity of a fractal object relative to its embedding space [45]. Generally, the relationship between the number of coverings (e. g. boxes in the box-counting method) $N(a)$ and the scaling factor $a$ (e. g. box length) is expressed as

$$N(a) \propto a^{-D} \tag{1}$$

where $D$ is the fractal dimension that can take positive non-integer values. Equation (1) represents the scaling law, which describes self-similarity of a fractal object using a single parameter. The fractal dimension $D$ was estimated by the box-counting method [46] using the fractal analysis MATLAB toolbox [47] by counting the number of boxes $N(a)$ with different side length $a$ as

$$D = \lim_{a \to 0} \frac{\ln N(a)}{\ln\left(\frac{1}{a}\right)}.$$ (2)

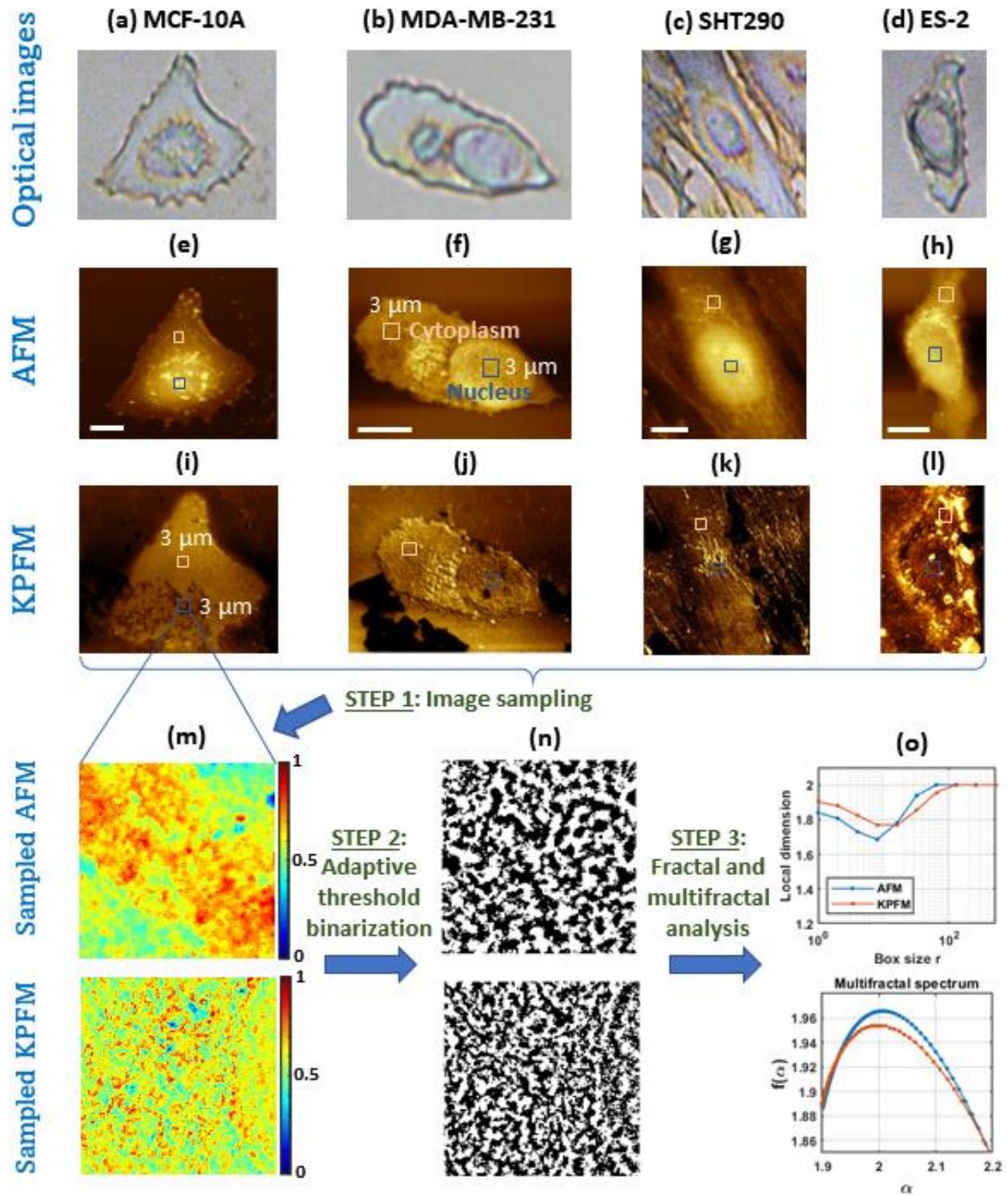

**Figure 1**. Illustration of the fractal and multifractal analysis procedure applied to AFM and KPFM images of normal (MCF-10a, SHT290) and cancer (MDA-MB-231, ES-2) cells. Scale bar in I – (h) is 10 μm.

However, a single parameter describing the monofractal geometry is limited in the ability to completely differentiate the complex scaling behaviors of many real-life objects [48-51]. This is attributed to the fact that the fractal dimension $D$ only characterizes the average fractality of multifractal objects via a simple scaling law [45, 52].

**Multifractal analysis**

Multifractal analysis addresses the limitations of the fractal analysis by providing a local description of the complex scaling behavior of biomedical images, represented as a spectrum of singularity exponents $f(\alpha)$ [31]. In the case of the binarized AFM/KPFM images, the local mass density $P_i(a)$ of the box centered at image location $x_i$ with side length $a$ is

$$P_i(a) = \frac{N_i(a)}{N},$$ (3)

where $N_i(a)$ and $N$ are the number of black pixels in box $i$ of size $a$ and the total number of pixels in box $i$, respectively. The scaling of $P_i(a)$ follows the power law, $P_i(a) = a^{\alpha_i}$, where $\alpha_i \equiv \alpha(x_i)$ is the singularity exponent describing the local scaling (local fractal behavior) centered at location $x_i$ of the $i^{th}$ box, which can be estimated as

$$\alpha_i = \lim_{a \to 0^+} \frac{\ln P_i(a)}{\ln(a)}.$$ (4)

The multifractal spectrum $f(\alpha)$ is defined as the fractal dimension of the set of locations $x$ such that $\alpha(x) = \alpha$. The $f(\alpha)$ spectrum represents a statistical distribution of the singularity exponents $\alpha_i$ which characterizes the local singularity strength or multifractality of the image. It is practically estimated using the Legendre transformation as

$$f(\alpha) = q\alpha - \tau(q), \ \alpha(q) = \frac{d\tau(q)}{dq},$$ (5)

where $q$ is the moment and $\tau(q)$ is the mass or Holder exponent [53]. In the multifractal formalism, $D_q = \tau(q)/(q-1)$ is the generalized fractal dimension, for which $D_0$ ($q = 0$) is equivalent to the box-counting fractal dimension. Furthermore, the information dimension $D_1$ ($q = 1$) describes the change of information entropy with the box size [54], whereas the correlation dimension $D_2$ ($q = 2$) quantifies the correlation of fractal measures in the boxes [55]. In addition, the width of the singularity spectrum $\Delta\alpha(Q) = \alpha_{max}(Q) - \alpha_{min}(Q)$, where $\alpha_{max} = \max\{\alpha(q), q \in [-Q, Q]\}$ and $\alpha_{min} = \min\{\alpha(q), q \in [-Q, Q]\}$, is the measure of multifractality [43]. The variation of the generalized fractal dimension $D_q$ over a range of $q$, known as the Rényi spectrum, characterizes the multifractal behavior of images, with the broader range of the sigmoidal curve corresponding to the more heterogeneous scaling [43, 56, 57]. The multifractal analysis was performed using the a MATLAB toolbox [49].

### *2.4. Statistical analysis and hypothesis testing*

To determine if the differences of the estimated multifractal parameters ($D_0$, $D_1$, $D_2$, and $\Delta\alpha(4)$) between the MCF-10A (normal) vs MDA-MB-231 (cancer) cell lines and SHT290 (normal) vs ES-2 (cancer) cell lines are statistically significant, two-sample $t$-test was performed on the measurements of the multifractal parameters, which were obtained from both AFM and KPFM images. The null hypothesis states that the random samples are from normal distributions with equal means and equal but unknown variances, and the alternative hypothesis states that the samples come from populations with unequal means. We used $n_1 = 4$ and $n_2 = 8$ number of AFM images for MCF-10A vs MDA-MB-231, and $n_1 = 5$ and $n_2 = 3$ for SHT290 vs ES-2 cells. We used $n_1 = 2$ and $n_2 = 11$ number of KPFM images for MCF-10A vs MDA-MB-231, and $n_1 = 3$ and $n_2 = 5$ for SHT290 vs ES-2 cells.

## 3. RESULTS

Typical representative examples of the AFM and KPFM images of the four cell lines and their corresponding binarized images, local fractal dimensions, and multifractal spectra using the adaptive and median value thresholding methods are shown in supplementary Figs. S2 and S3 (see Supplementary Information). The thresholding values of the image binarization cut-off were estimated based on the median value of all pixels for the median method and the local mean intensity of the pixel neighborhood for the adaptive method, as described in the Methods section. The adaptive method results were more consistent and showed better statistical significance in distinguishing normal and cancer cells as described below. This may be due to the presence of the occasional random dust particles on the sample that have a large height in AFM images, affecting the median height values. This effect is minimized in the adaptive method. Therefore, we selected the adaptive method below to demonstrate the results. We included the median method results in the supplementary Figs. S2 and S3 for comparison. While both methods showed similar results for most of the cells, some of the data showed significant differences due to the random fluctuations, which were especially pronounced in the AFM height images. For example, Figs. S2m and S2q show similar fractal and multifractal AFM plots for an MCF-10A cell, while Figs. S2n and S2r show large differences for an MDA-MB-231 cell. However, the corresponding KPFM signals from the same cells showed similar plots in Figs. S2o, S2p, S2s and S2t. Similar trends are observed for the ovarian cells in supplementary Fig. S3, which showed large differences in the AFM and consistent results in the KPFM plots. This demonstrates the advantage of the KPFM method for cancer detection compared to AFM. This is because AFM is more sensitive to ambient dust contamination that introduces large random height fluctuations, while KPFM measures bioelectric signals, which are less sensitive to random height fluctuations from the occasional uncharged particles.

Supplementary Fig. S4 shows typical representative AFM and KPFM profiles that were obtained by averaging 5 lines of the 3×3 μm$^2$ scan areas used in the multifractal analysis. The profiles show the amplitude and correlation between fluctuations of height (AFM) and surface potential (KPFM) of normal and cancer cells. The linear correlation coefficients between the AFM and KPFM profiles of MCF-10A, MDA-MB-231, SHT290, and ES-2 cells are 0.696, 0.434, 0.038, and 0.362, respectively. The results show larger spatial fluctuations of surface potential of cancer cells compared to normal cells, characterized by fractal and multifractal analyses.

### *Fractal analysis*

First, we present the results of fractal analysis of AFM and KPFM images using the box-counting method with adaptive thresholding. Figure 2 shows the dependence of the local fractal dimension $D_f$ as a function of the box size $r$ that was obtained from the slope of the tangent of $N(r)$ at each value of $r$ estimated from the box-counting algorithm ($D_f = -d \ln N(r) / d \ln r$) [58]. Thick solid lines correspond to the mean values, while thin dashed lines show individual cell data. The supplementary Fig. S5 shows the corresponding results using the median threshold analysis. These figures show similar average trends as the examples of the individual cells described above. KPFM shows more consistent results compared to AFM due to the smaller sensitivity to random particles.

The main insight from the fractal analysis in Fig. 2 is the observation that the local fractal dimension is not constant over the range of box sizes and varies for the different cell lines. This leads to the conclusion that both AFM and KPFM images cannot be described as simple fractal objects. They have a more complex organization that depends on the scaling range. This requires the multifractal analysis shown below. However, fractal analysis can still be used to distinguish normal and cancer cells. It can show general trends in the variations of the local height and surface potential. For example, Fig. 2a shows smaller cell-to-cell variations in AFM local dimensions for MDA-MB-231 cells (red dashed lines) compared to MCF-10A cells (blue dashed lines). This indicates that MDA-MB-231 cell topography is closer to being fractal than MCF-10A. MDA-MB-231 cell height variations may be approximately described by a fractal dimension of ~ 1.9, while the fractal dimension of MCF-10A cells varies widely within ~ 1.5 – 2 range. KPFM images show similar trends with different fractal dimensions. A smaller average range of local fractal dimensions of ~ 1.6 – 2 for MDA-MB-231 cells (thick solid red line in Fig. 2b) indicates a more fractal-like description compared to the broader average range of ~ 1.5 – 2 for MCF-10A cells (thick solid blue line in Fig. 2b). Overall, both breast and ovarian cell lines exhibited a loss of multifractality upon cancerous transformation that is confirmed by the multifractal analysis below. Also, KPFM data for MDA-MB-231 cells (red dashed lines in Fig. 2b) show larger cell-to-cell variations compared to MCF-10A cells. This reflects a larger heterogeneity of the surface potential of cancer cells compared to normal cells.

The AFM fractal analysis of the ovarian cells in Fig. 2c shows similar trends, which are unable to distinguish between the normal and cancer cells. On the other hand, KPFM shows a better distinguishability in Fig. 2d. Overall, although fractal analysis can be used to distinguish between the normal and cancer cells, it provides limited insights into the multiscale organization of the morphology and surface potential.

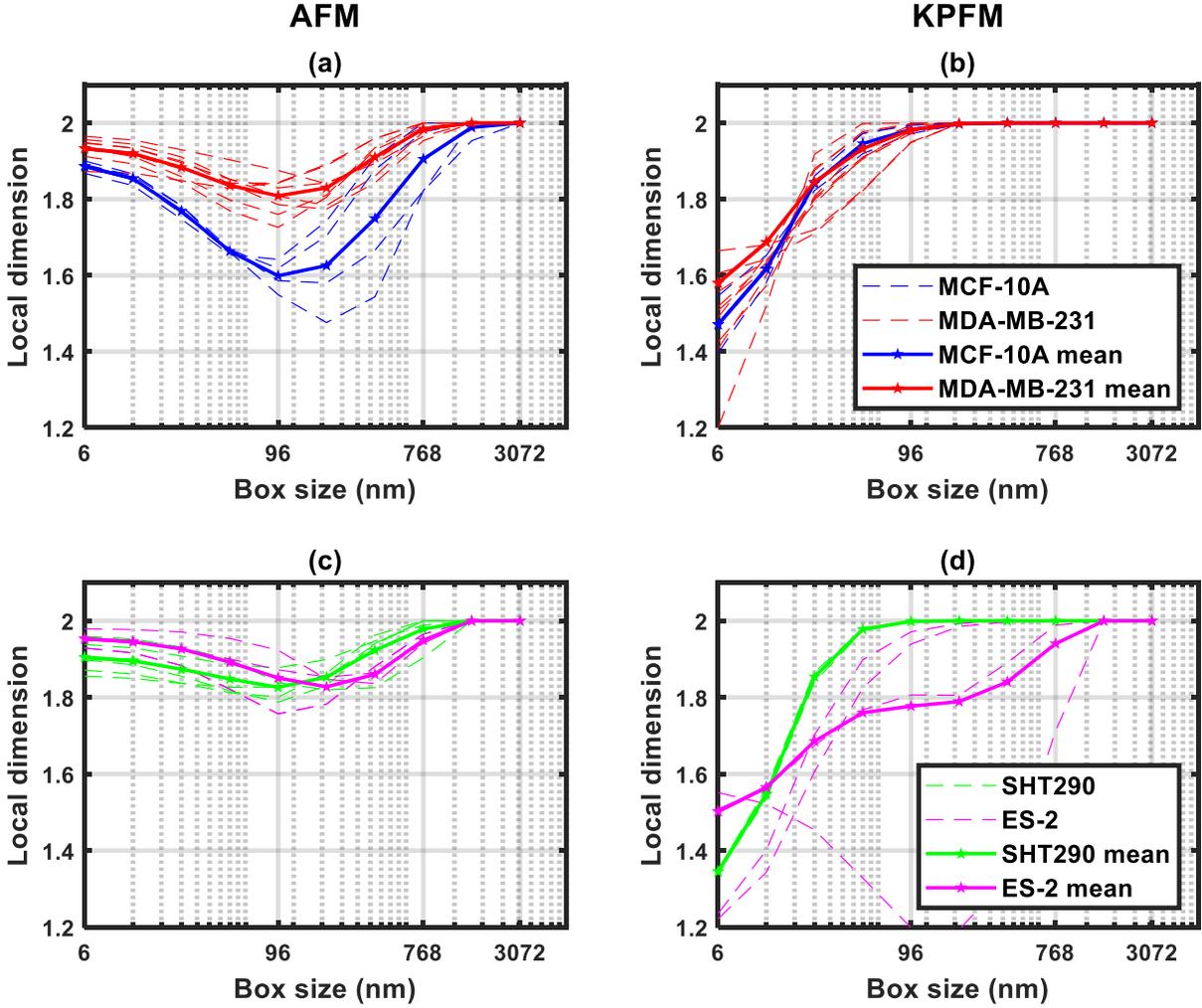

**Figure 2**. Fractal analysis of AFM and KPFM images using the box-counting method with adaptive thresholding. Dependence of the local fractal dimension $D_f$ as a function of the box size for individual cells (thin dashed lines) and mean data (thick solid lines).

*Multifractal analysis*

Next, we performed multifractal analysis of the same AFM and KPFM images of the four cell lines that were used for the fractal analysis above. Figure 3 shows the Rényi spectra $D_q$, multifractal spectra $f(\alpha)$, and singularity width $\Delta\alpha$ obtained using the adaptive method. Thick solid lines show the averages over all areas of the cells including both cytoplasm and nucleus. Thin dashed lines show individual areas. The corresponding results using the median method are shown in supplementary Fig. S6.

The Rényi spectra in Figs. 3a – 3d show significant differences between the normal and cancer cells at $D_1$ and $D_2$ values (vertical dashed lines) in KPFM compared to AFM images. In these plots, positive $q$ values accentuate denser regions, and negative $q$ accentuate the less dense regions. On the contrary, the Rényi spectra show larger differences between the normal and cancer cells at $D_0$ for AFM compared to KPFM. These results are summarized in Tables 1 and 2 below. The median method did not give satisfactory results for AFM images (see supplementary Figs. S6a and S6c). However, it showed satisfactory results for the KPFM images (supplementary Figs. S6b and S6d), confirming the conclusion from the fractal analysis that surface potential is less sensitive than topography to random height variations.

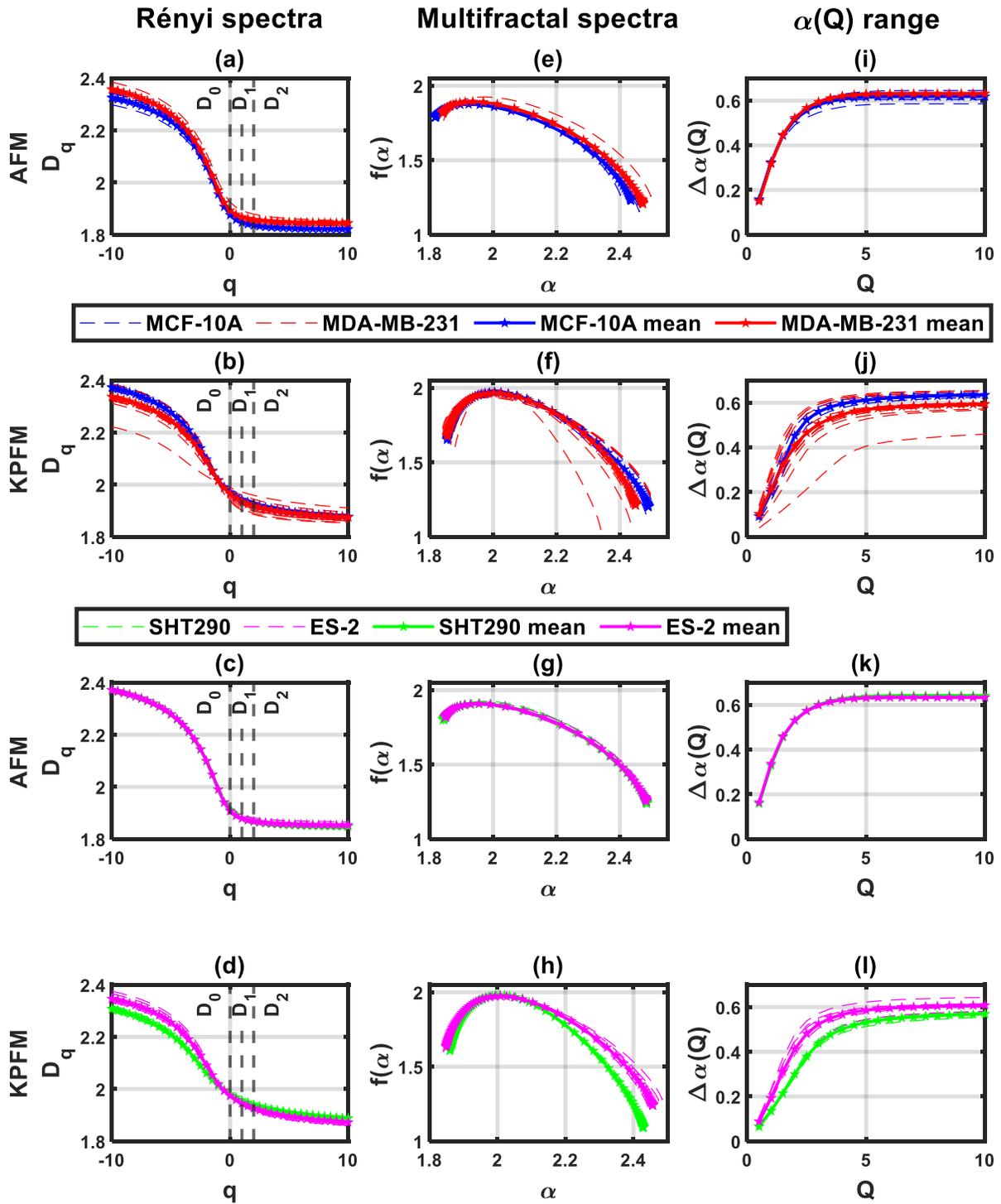

**Figure 3**. Rényi spectra (a – d), multifractal spectra (e - h), and singularity exponent range (i - l) obtained from multifractal analysis of adaptive AFM and KPFM images of normal and cancer cells for individual areas (dashed lines) and averages (solid lines). Vertical dashed lines mark the $D_0$, $D_1$ and $D_2$ values.

The multifractal spectra in Figs. 3e – 3h provide direct information about the multifractality of AFM and KPFM images. The concave shape of the spectra indicates multifractality, which is quantified by the width $\Delta\alpha$. The wider spectra with the larger $\Delta\alpha$ correspond to higher multifractality. Figure 3e shows multifractal spectra for AFM images of breast cells that have similar width $\Delta\alpha$ but different $D_0$, i. e. the MDA-MB-231 spectra (red lines) are shifted to higher $\alpha$ values. The corresponding KPFM spectra in Fig. 3f show similar shapes with similar $D_0$ and $\Delta\alpha$ parameters. The KPFM spectra in Fig. 3f cannot distinguish between the normal and cancer cells, but they quantify multifractality and provide further insights to support the results of fractal analysis in Fig. 2.

Multifractal analysis of the ovarian cells showed qualitatively different results. While no significant difference was observed in the AFM spectra (Fig. 3g), there was a significant difference in the width $\Delta\alpha$ of ES-2 cells (magenta lines) having larger $\Delta\alpha$ than SHT290 (green lines) as shown in Fig. 3h. Also, AFM spectra of both ES-2 and SHT290 cells showed small cell-to-cell variations in the whole $\alpha$ range in Fig. 3g, while KPFM spectra showed larger variations for ES-2 compared to SHT290 at large $\alpha$ values (Fig. 3h). This indicates heterogeneity of the cancer cell surface potential despite the small variations in cell morphology.

We compared the adaptive and median image binarization methods for fractal and multifractal analyses. The adaptive method showed standard concave $f(\alpha)$ curves for both AFM and KPFM images, while the median method failed. The adaptive method showed statistically significant differences between the KPFM multifractality for the ovarian cancer cells but the corresponding analysis using the median method showed large errors and non-concave $f(\alpha)$ curve shapes for AFM.

The main result in Figs. 3g and 3h is that the surface potential of cancer cells is more multifractal compared to normal cells. Multifractality is more pronounced at large values of the momentum $Q$ as shown in the Methods sections. Figures 3i – 3l show the $\Delta\alpha(Q)$ plots to better visualize multifractality at different $Q$ values. Figs. 3k and 3l show a drastic difference between the multifractality changes of AFM and KPFM images of the ovarian cells. It supports the conclusion that $\Delta\alpha$ of surface potential can be used as a cancer biomarker.

Figure 4 shows the comparison of the parameters $D_0$, $D_1$, $D_2$, and $\Delta\alpha(4)$ from the cytoplasm, nucleus, and combined regions using adaptive method. The corresponding results using the median method are shown in supplementary Fig. S7. Figs. 4a, 4c and 4d show that $D_0$, $D_1$, and $D_2$ of the AFM images of MCF-10A are lower than that of MDA-MB-231 ($D_0$ at $p < 0.05$; $D_1$ at $p < 0.01$; $D_2$ at $p < 0.001$; see Table 1). This indicates the increase of the morphological local fractal dimension and heterogeneity (disorder) in breast cancer cells. Fig. 4b shows no significant difference in the mean value of $\Delta\alpha(4)$ of the AFM images of the four cell types but MCF-10A cells showed the largest variance of $\Delta\alpha(4)$.

Figure 4 shows the opposite behavior of the AFM and KPFM multifractal parameters. In particular, Figs. 4e, 4g and 4h show that $D_0, D_1$, and $D_2$ of KPFM images of SHT290 are larger than those of ES-2 ($p < 0.1$; Table 2), which indicates the lower local fractal dimension of surface potential in ovarian cancer cells compared to normal endometrial cells. However, the $D_0, D_1$, and $D_2$ parameters have larger variations for ES-2 compared to SHT290 cells, possibly reflecting a larger heterogeneity of the cancer cells. Fig. 4f shows a larger mean value of $\Delta\alpha(4)$ for ES-2 compared to SHT290 ($p < 0.05$), indicating larger multifractality of the ovarian cancer cell surface potential. No significant difference between the mean values of the surface potential $\Delta\alpha(4)$ were found for MCF-10A and MDA-MB-231 cells. However, the MDA-MB-231 cells showed larger variations of $\Delta\alpha(4)$.

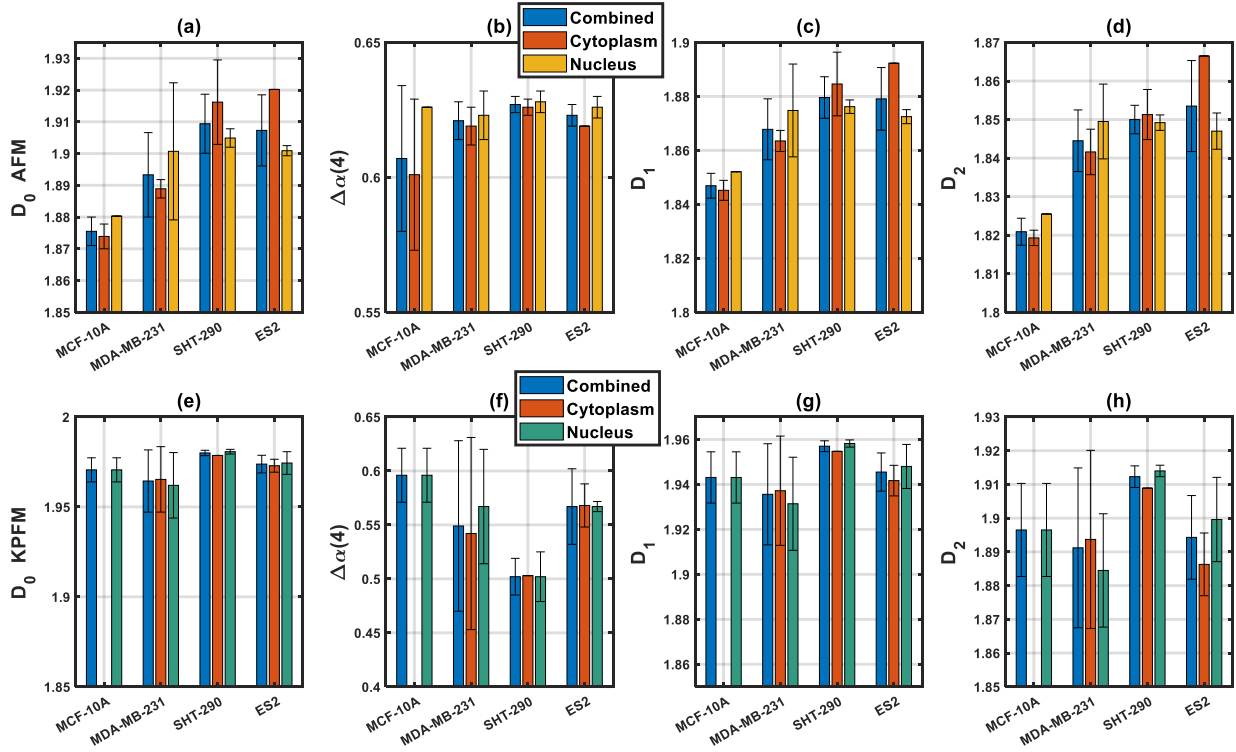

**Figure 4**. Comparison of the multifractal parameters $D_0$, $D_1$, $D_2$, and $\Delta\alpha(4)$ from the combined and separate cytoplasm and nucleus regions using adaptive method.

The hypothesis testing results for the differences in the multifractal parameters of AFM and KPFM images using the adaptive method are shown in Tables 1 and 2. The hypothesis testing for the mean difference in the multifractal parameters of AFM images revealed the significant differences in $D_0$, $D_1$, and $D_2$ between the MCF-10A and MDA-MB-231 cell lines at the significance level of 0.05, but no significant difference in $\Delta\alpha(4)$ was found (Table 1). However, for the ovarian cell lines, no significant differences in the four multifractal parameters of the AFM images of SHT290 and ES-2 cell lines were identified.

**Table 1**. Hypothesis testing for the difference in the multifractal parameters (box counting dimension $D_0$, entropy dimension $D_1$, correlation dimension $D_2$, and singularity exponent range $\Delta\alpha(4)$) of AFM images using the adaptive method.

| Hypothesis testing | Sample pair | $\overline{X}_1 \pm s_1$ | $\overline{X}_2 \pm s_2$ | $p$-value |
|---|---|---|---|---|
| Mean difference in $D_0$ | MCF-10A vs MDA-MB-231 | $1.876 \pm 0.005$ | $1.894 \pm 0.013$ | 0.0286 *[a] |
| | SHT290 vs ES-2 | $1.909 \pm 0.009$ | $1.907 \pm 0.011$ | 0.7801 |
| Mean difference in $D_1$ | MCF-10A vs MDA-MB-231 | $1.847 \pm 0.005$ | $1.868 \pm 0.011$ | 0.0059** |
| | SHT290 vs ES-2 | $1.880 \pm 0.008$ | $1.880 \pm 0.012$ | 0.9429 |
| Mean difference in $D_2$ | MCF-10A vs MDA-MB-231 | $1.821 \pm 0.004$ | $1.845 \pm 0.008$ | 0.0002*** |
| | SHT290 vs ES-2 | $1.850 \pm 0.004$ | $1.854 \pm 0.012$ | 0.5434 |
| Mean difference in $\Delta\alpha(4)$ | MCF-10A vs MDA-MB-231 | $0.607 \pm 0.027$ | $0.621 \pm 0.007$ | 0.1793 |
| | SHT290 vs ES-2 | $0.627 \pm 0.003$ | $0.623 \pm 0.004$ | 0.1615 |

[a] Significance codes: $p < 0.001$ '***', $p < 0.01$ '**', $p < 0.05$ '*'

In contrast to the hypothesis testing results of the AFM images, the hypothesis tests for the KPFM images mean difference in $D_0$, $D_1$, and $D_2$ showed the significant differences between the SHT290 and ES-2 cell lines at the significance level of 0.1 and the significant difference in $\Delta\alpha(4)$ at the level of 0.05, but there were no significant differences between MCF-10A and MDA-MB-231 cell lines in terms of the four multifractal parameters (Table 2).

**Table 2**. Hypothesis testing for the difference in the multifractal parameters (box counting dimension $D_0$, entropy dimension $D_1$, correlation dimension $D_2$, and singularity exponent range $\Delta\alpha(4)$) of KPFM images using the adaptive method.

| Hypothesis testing | Sample pair | $\overline{X}_1 \pm s_1$ | $\overline{X}_2 \pm s_2$ | $p$-value [a] |
|---|---|---|---|---|
| Mean difference in $D_0$ | MCF-10A vs MDA-MB-231 | $1.971 \pm 0.007$ | $1.964 \pm 0.017$ | 0.6354 |
| | SHT290 vs ES-2 | $1.980 \pm 0.002$ | $1.974 \pm 0.005$ | 0.0821 · |
| Mean difference in $D_1$ | MCF-10A vs MDA-MB-231 | $1.943 \pm 0.011$ | $1.936 \pm 0.023$ | 0.6645 |
| | SHT290 vs ES-2 | $1.957 \pm 0.002$ | $1.946 \pm 0.009$ | 0.0656 · |
| Mean difference in $D_2$ | MCF-10A vs MDA-MB-231 | $1.897 \pm 0.014$ | $1.845 \pm 0.008$ | 0.7685 |
| | SHT290 vs ES-2 | $1.912 \pm 0.003$ | $1.894 \pm 0.012$ | 0.0536 · |
| Mean difference in $\Delta\alpha(4)$ | MCF-10A vs MDA-MB-231 | $0.596 \pm 0.025$ | $0.549 \pm 0.079$ | 0.4339 |
| | SHT290 vs ES-2 | $0.502 \pm 0.017$ | $0.567 \pm 0.035$ | 0.0256* |

[a] Significance codes: $p < 0.05$ '*', $p < 0.1$ '·'

# 4. DISCUSSION

The observed differences between the morphological and surface potential measurements can be explained by the molecular structural and bioelectric changes of the plasma membrane. Our results suggest that AFM and KPFM are complementary imaging techniques that reflect different molecular mechanisms of tumorigenesis, such as the expression of membrane surface proteins, activity of ion channels and cytoskeletal reorganization. In the case of ovarian cancer cells (*i.e.*, normal-immortalized SHT290 and tumorigenic ES-2), KPFM provides significant differences in multifractality, while AFM does not. These observations agree with the molecular model of ovarian cancer based on the overexpression of certain proteins such as the protein kinase C (PKC) family [59, 60]. PKC activation by $Ca^{2+}$ may reflect the change in the surface potential. Although binding of PKC to the inner side of the plasma membrane does not affect the membrane outer surface morphology [60]. The previous study showed an overexpression of PKC proteins in ES-2 cells compared to SHT290 cells [60], it utilizes Ca2+ in a complex signaling cascade involved in cancer progression. Similar importance of PKC proteins was previously shown in breast cancer [61-63]. Overexpression of PKC-ξ protein was previously correlated with the invasiveness in MDA-MB-231 cells [63]. In addition, the PKC signaling pathway effects cytoskeleton reorganization by regulating the actin polymerization, lipid metabolism and $Ca^{2+}$ [64]. These and other morphological changes may be responsible for the differences that we observed in the AFM fractality of breast cells.

Previous studies reported a depolarization of the surface potential of cancer cells and tissues [5, 6]. Although ovarian ES-2/SHT290 cells exhibited a larger depolarization compared to breast MCF-10A/MDA-MB-231, of note is the higher variance among both tissue types in the cancerous cell lines compared to their normal-immortalized counterparts. The smaller size of the ovarian (~10-15 μm) compared to breast (~20-30 μm) and the small sample size may be the reason that our results on multifractality did not show a significant difference between the MCF-10A and MDA-MB-231 cells. Based on the larger difference in surface potential, we predict that the multifractal method may also work well on pancreatic, hepatic, and rectal cancer.

Note that our imaging was performed under steady-state conditions, in which the sub-millisecond temporal fluctuations of surface potential were averaged out. Therefore, our work is focused on spatial fluctuations. Recent work on temporal dynamics of the surface potential revealed larger fluctuations in MDA-MB-231 cells compared to MCF-10A cells [65]. Although potential fluctuations in cancer cells were inhibited by the voltage-gated Na channel and $Ca^{2+}$ - activated K channel blockers, which indicates the possible role of the ion channels in modifying the membrane morphology that we measured in our AFM experiments. However, the previous measurements in [65] were performed using voltage-sensitive dyes and were not label-free. Our KPFM measurements, on the other hand, are label-free and are minimally invasive.

The comparison between the nucleus and cytoplasm areas did not show any significant differences in both AFM and KPFM signals, indicating that the cell migration mechanisms of cytoskeletal reorganization do not contribute to the observed multifractality of surface potential. For example, $Ca^{2+}$ flickering was previously found to be predominant in lamellipodia at migrating cell edges [66]. High calcium microdomains were observed that could contribute to the spatial organization of surface potential and polarity between the leading and trailing edges. However, our study included both edges and showed no significant differences, ruling out the calcium-based migration mechanism.

To the best of our knowledge, our work shows multifractality in surface potential of biological cells for the first time. Fractal and multifractal potentials may lead to large fluctuations of local electric fields, such as electric hot spots in nanoplasmonics systems used for sensing applications [67-69]. Large electric fields in biological cell systems may be attributed to fractal aggregation of charged proteins [70] accompanied by complex heterogeneous potential landscapes [71, 72].

## 5. CONCLUSION

In summary, we measured the surface potential of normal and cancer cells with nanoscale resolution. We used fractal and multifractal analysis to characterize the time-averaged spatial potential fluctuations and observed changes in multifractality of ES-2 ovarian cancer cells compared to control SHT290 normal endometrial cells. The ovarian cancer cells had a more multifractal surface potential compared to the normal cells. While AFM morphology failed to distinguish between normal and cancer cells, we observed multifractality in the surface potential of breast cancer cells, which, however, did not show a significant difference from the normal breast cells. We observed significant differences in the AFM morphological fractal dimension of these non-malignant and malignant breast cells. Based on the comparison between the morphological and surface potential images of normal and cancer cells, we conclude that the combination of these two imaging techniques improves cancer detection by providing complementary information. Cell functions such as immune signaling, metabolism and apoptosis are dysregulated during tumor growth, resulting in catastrophic physical and bioelectric changes that can be characterized by the multifractality of the AFM and KPFM images. The characterized morphological and surface potential images will facilitate our previous works on probabilistic modeling of disorder pathogenesis [73-75] by adding more details from cell and tissue levels. Multifractality in surface potential can also be used as a novel biomarker for novel drug delivery methods based on bioelectric activation.

## DECLARATION OF COMPETING INTERESTS

The authors declare that they have no known competing financial interests or personal relationships that could have appeared to influence the work reported in this paper.

# Supplementary Information

# Multifractality in Surface Potential for Cancer Diagnosis

Phat K. Huynh [1], Dang Nguyen [2], Grace Binder [3], Sharad Ambardar [2], Trung Q. Le [1,2], and Dmitri V. Voronine [2,4*]

[1] Department of Industrial and Management Systems Engineering, University of South Florida, Tampa, FL, 33620

[2] Department of Medical Engineering, University of South Florida, Tampa, FL, 33620

[3] Department of Chemistry, University of South Florida, Tampa, FL, 33620

[4] Department of Physics, University of South Florida, Tampa, FL, 33620

* Corresponding author: dmitri.voronine@gmail.com

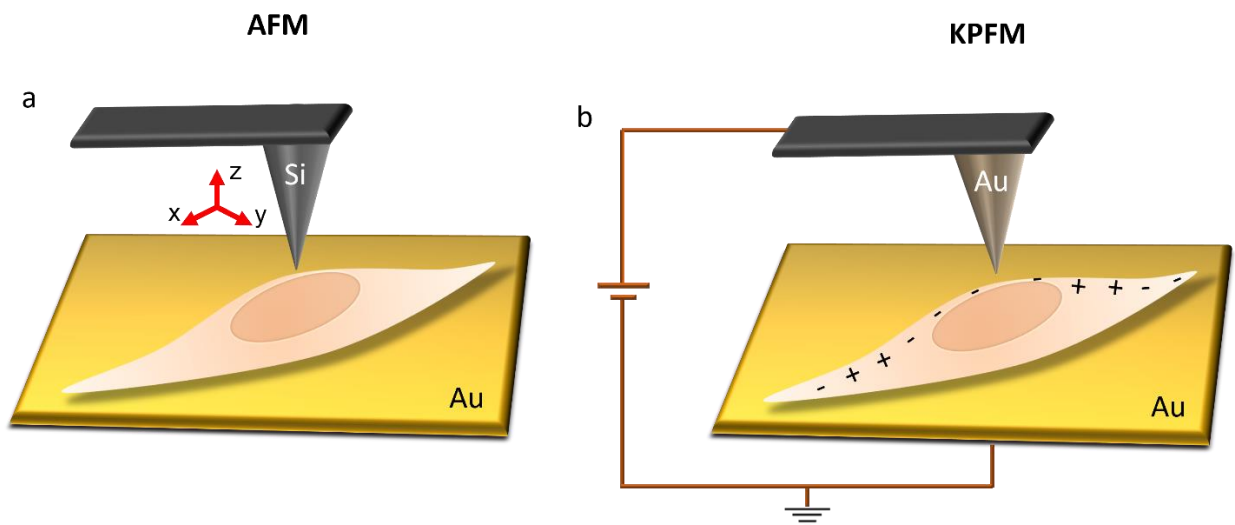

**Figure S1**. Schematic diagrams of the morphological AFM (a) and surface potential KPFM (b) imaging of cells.

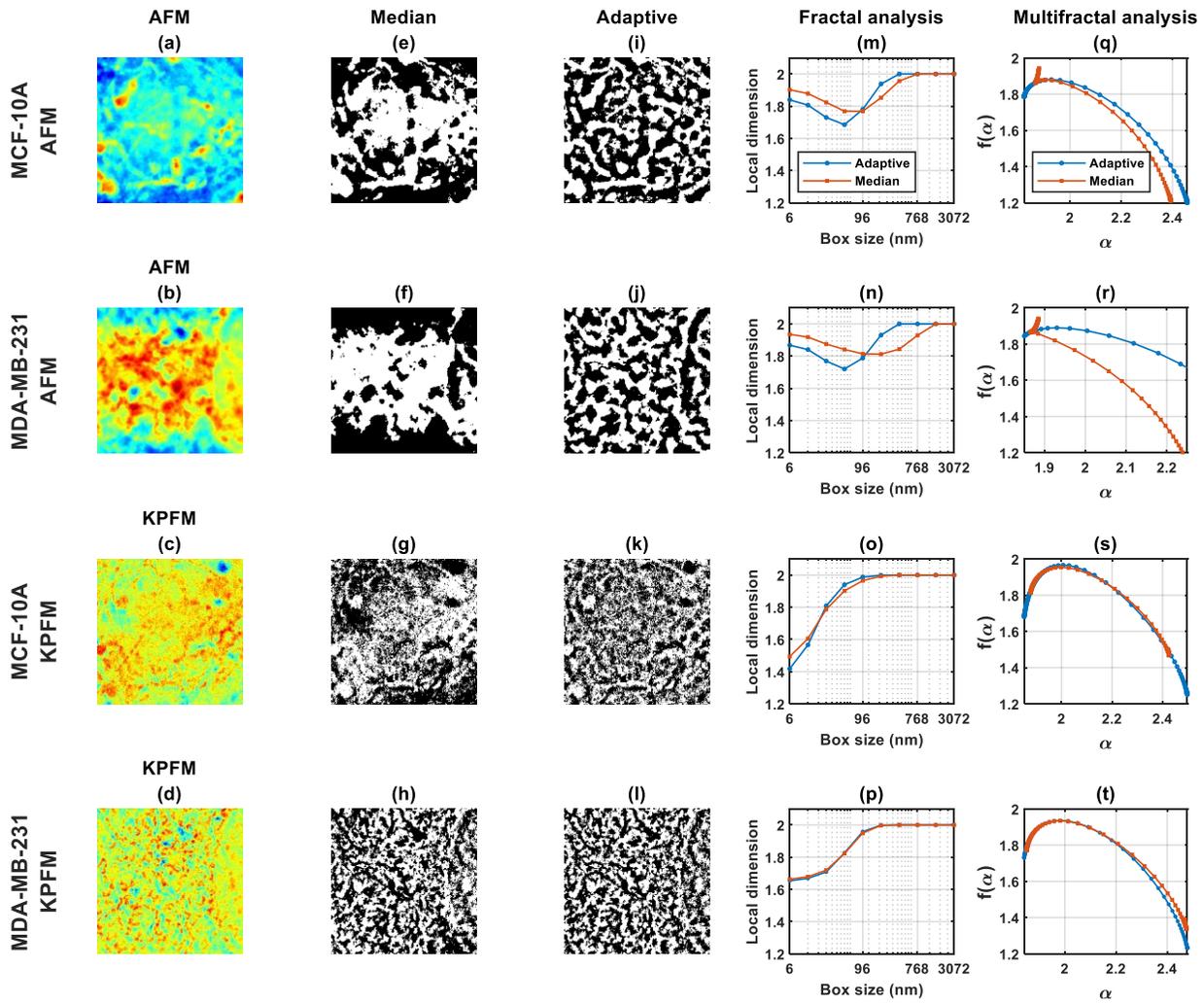

**Figure S2**. Typical representative examples of the AFM (a,b) and KPFM (c,d) images of the breast cell lines and their corresponding binarized images (e - l), local fractal dimensions (m - p), and multifractal spectra (q - t) using the adaptive and median value thresholding methods.

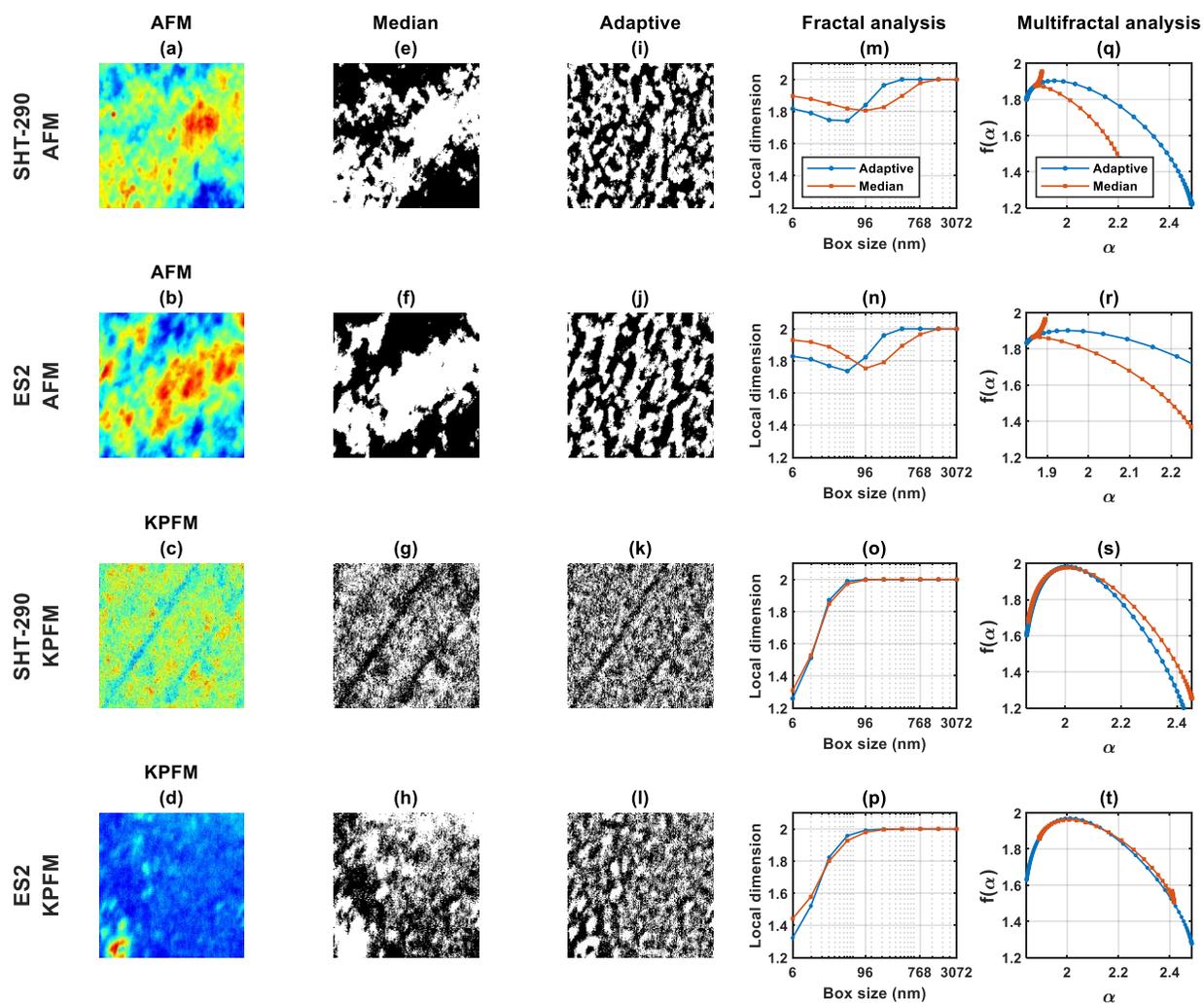

**Figure S3**. Typical representative examples of the AFM (a,b) and KPFM (c,d) images of the ovarian cell lines and their corresponding binarized images (e - l), local fractal dimensions (m - p), and multifractal spectra (q - t) using the adaptive and median value thresholding methods.

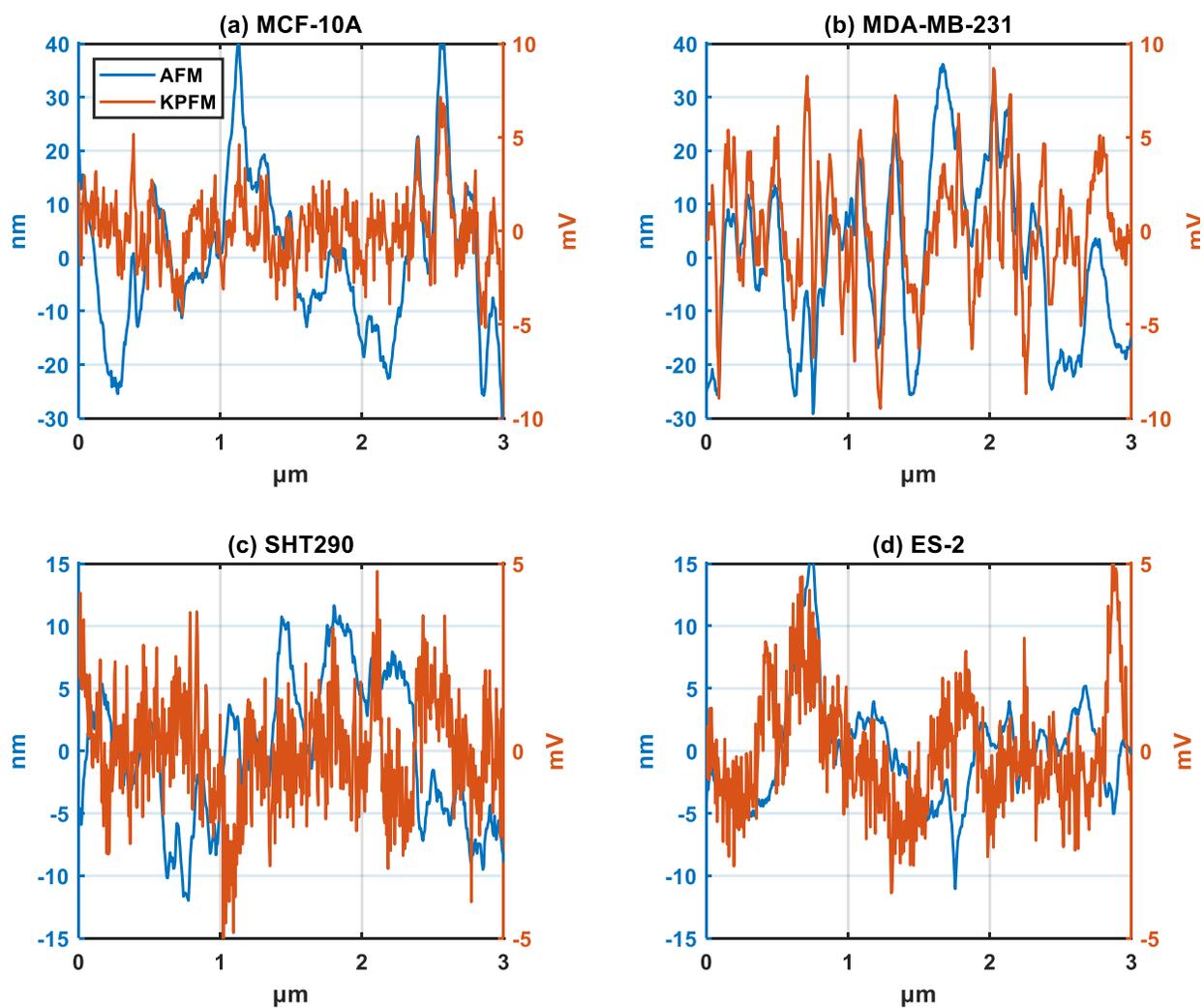

**Figure S4**. Typical representative AFM and KPFM profiles obtained from the 3×3 µm² scan areas used in the multifractal analysis.

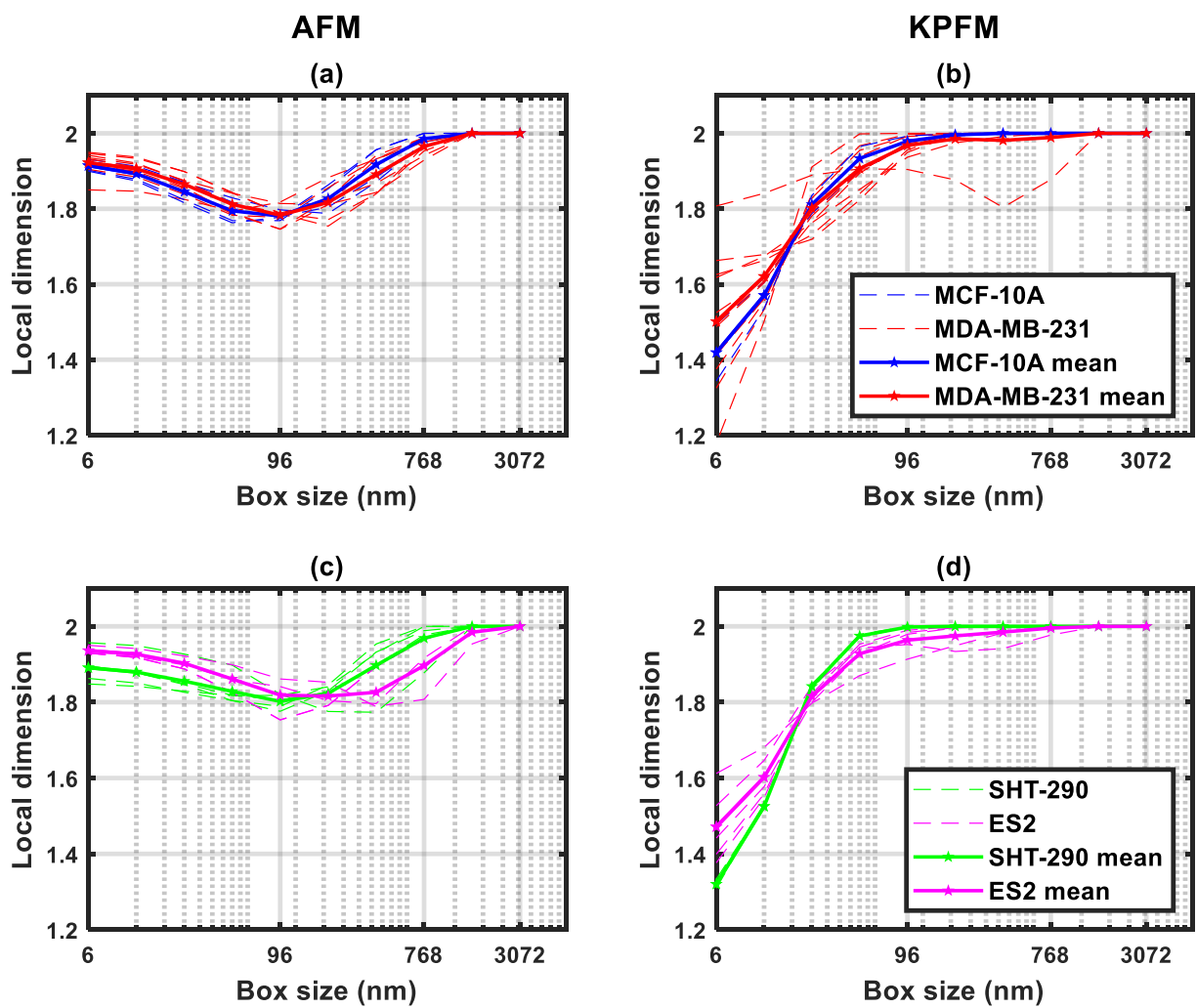

**Figure S5**. Fractal analysis of AFM and KPFM images using the box-counting method with median thresholding. Dependence of the local fractal dimension $D_f$ as a function of the box size for individual cells (thin dashed lines) and mean data (thick solid lines).

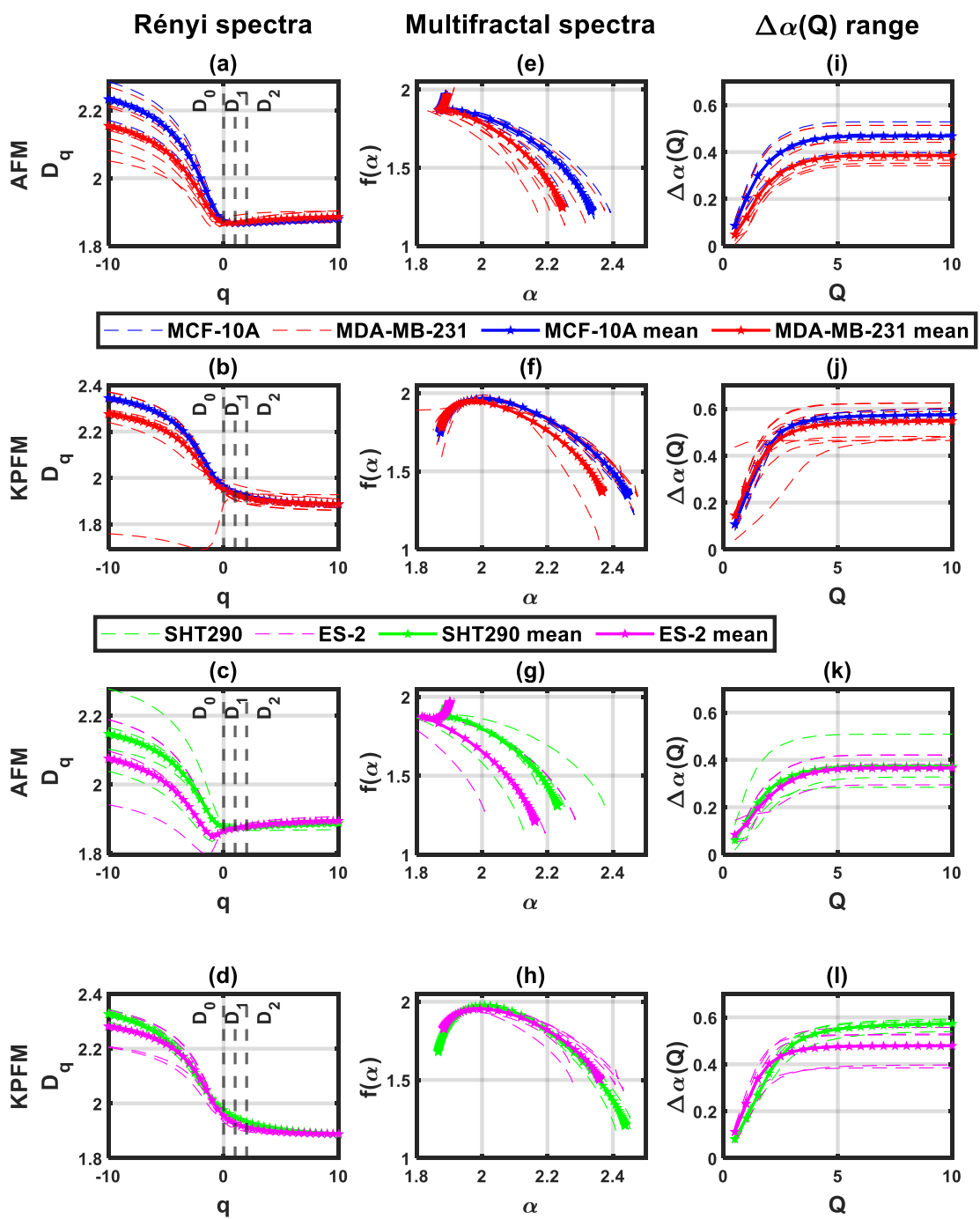

**Figure S6**. Rényi spectra (a – d), multifractal spectra (e – h), and singularity exponent range (i - l) obtained from multifractal analysis of median AFM and KPFM images of normal and cancer cells for individual areas (dashed lines) and averages (solid lines). Vertical dashed lines mark the $D_0$, $D_1$ and $D_2$ values.

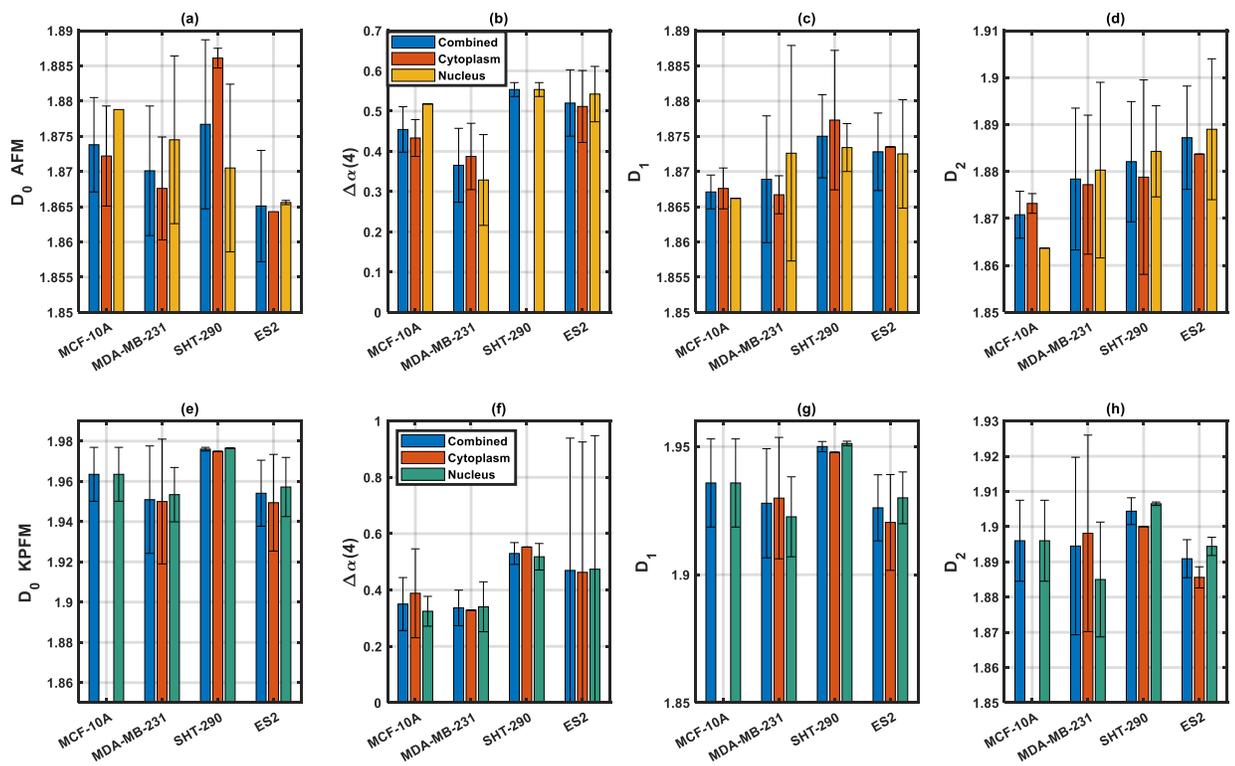

**Figure S7**. *Comparison of the multifractal parameters $D_0$, $D_1$, $D_2$, and $\Delta\alpha(4)$ from the combined and separate cytoplasm and nucleus regions using the median method.*